Deep Learning models for benign and malign Ocular Tumor Growth Estimation

Mayank Goswami

Divyadrishti Imaging Laboratory, Department of Physics, Indian Institute of Technology Roorkee, Roorkee, India
*mayank.goswami@ph.iitr.ac.in

## Abstract

Relatively abundant availability of medical imaging data has provided significant support in the development and testing of Neural Network based image processing methods. Clinicians often face issues in selecting suitable image processing algorithm for medical imaging data. A strategy for the selection of a proper model is presented here. The training data set comprises optical coherence tomography (OCT) and angiography (OCT-A) images of 50 mice eyes with more than 100 days follow-up. The data contains images from treated and untreated mouse eyes.

Four deep learning variants are tested for automatic (a) differentiation of tumor region with healthy retinal layer and (b) segmentation of 3D ocular tumor volumes. Exhaustive sensitivity analysis of deep learning models is performed with respect to the number of training and testing images using ~~8~~ eight performance indices to study accuracy, reliability/reproducibility, and speed. U-net with UVgg16 is best for malign tumor data set with treatment (having considerable variation) and U-net with Inception backbone for benign tumor data (with minor variation). Loss value and root mean square error (R.M.S.E.) are found most and least sensitive performance indices, respectively. The performance (via indices) is found to be exponentially improving regarding a number of training images. The segmented OCT-Angiography data shows that neovascularization drives the tumor volume. Image analysis shows that photodynamic imaging-assisted tumor treatment protocol is transforming an aggressively growing tumor into a cyst.

An empirical expression is obtained to help medical professionals ~~to~~ choose a particular model given the number of images and types of characteristics.

We recommend that the presented exercise should be taken as standard practice before employing a particular deep learning model for biomedical image analysis.

## 1. Introduction

Primary intraocular cancer is a rare occurrence.
The majority of adult patients (age > 15 years) suffer uveal melanoma. Young children (median age 5 years) suffer retinoblastoma, orbital rhabdomyosarcoma, medulloepithelioma, and ocular non-Hodgkin's lymphoma. We refer (Arora et al., 2009; Broaddus et al.; 2009; Jain et al., 2019; Talkington and Durrett, 2015; "The Collaborative Ocular Melanoma Study (COMS) randomized trial of pre-enucleation radiation of large choroidal melanoma II: initial mortality findings COMS report no. 10," 1998; Verbraeken et al., 1997) for statistical studies, in brief. Uveal melanoma affects mostly the choroidal region with relatively less severity in the iris and ciliary body (Costache et al., 2013; Schoenfield, 2014). Retinoblastoma affects cone photoreceptors and can be diagnosed easily (Dimaras et al., 2015; Dimaras and Corson, 2019). The majority of retinoblastoma cases are reported in the Asia-pacific region (Jain et al., 2019).

Rhabdomyosarcoma, a rapidly growing tumor, occurs inside the orbital region of the eye with a rare incidence of metastasizing into lung, bone, and bone marrow with life-threatening conditions (Jurdy et al., 2013).

Intraocular medulloepithelioma appears as a ciliary body tumor with slow growth. It can also originate from the retina or optic nerve (Peshtani et al., 2014). Ocular lymphoma can originate from the retina, uvea, vitreous, Bruch's membrane, and optic nerve with overlapping symptoms with other infectious diseases (Tang et al., 2017). Its diagnosis still a challenging task for both ophthalmologists and pathologists.

The success rate of prognosis depends upon early detection in all the above types of eye cancers. Failure results in permanent blindness, removal of the eye (enucleation), and death in the worst-case scenario. The treatment path utilizes radiation therapy, cryotherapy, immunotherapy, chemotherapies, or combinations of more than one of these techniques ("The Collaborative Ocular Melanoma Study (COMS) randomized trial of pre-enucleation radiation of large choroidal melanoma II: initial mortality findings COMS report no. 10," 1998). Once diagnosed, treatment planning evolves according to the growth rate of the tumor and its response against the treatment. Patient's age and longest basal diameter of the melanoma have been reported to affect the prognosis to a statistically appreciable degree ("The Collaborative Ocular Melanoma Study (COMS) randomized trial of pre-enucleation radiation of large choroidal melanoma II: initial mortality findings COMS report no. 10," 1998). Power law, exponential, Gompertz, and Spratt's generalized logistic model are used to fit and predict the growth rate of the tumor using in vitro data (Talkington and Durrett, 2015). Surface growth for two neurological cancers was found to follow 2/3's power law. Optimized in-vivo imaging techniques are suggested to reduce the uncertainty affecting the estimation of doubling time, percentage increase per unit time, and specific growth rate (Mehrara et al., 2009). Noninvasive, *in-vivo*, 3D imaging techniques allow us to track the cancer progressions in its natural homeostatic environment (Iafrate and Fruhwirth, 2020). Ocular Ultrasound B-scan (USB) and Ultrasound biomicroscopy (U.B.M.) images depict boundaries of lens, choroid, retina, and sclera with a particular trade-off between axial and lateral resolution (Hau et al., 2015). These are preferred modalities to acquire images right behind the iris, including the ciliary body and lens, especially diagnosing anterior scleral inflammatory disease, glaucoma, and cataracts symptoms. These techniques lack the ability to differentiate between active and inactive tumor and interior iris liaisons (Dimaras and Corson, 2019). The acquisition mode needs a water-bath immersion with direct contact to the eye, longer image acquisition times, and the need for an experienced operator (Hau et al., 2015). Magnetic Resonance Imaging (M.R.I.) has been used as a data reconstruction techniques in cases where direct access to the interior region of the eye is blocked due to injury, cataract, imaging during the surgery, etc. (Materin et al., 2012). M.R.I. images can show differentiation morphologically between types of cancer (Malikova et al., 2016). Fundus imaging provides a quick assessment of visualization of *en-face* (2D front view) of the eye. Optical Coherence Tomography (OCT) and OCT-Angiography (OCT-A) are proven photographic/direct imaging modalities, providing clinicians high-resolution time-dependent qualitative as well quantitative information (Fujimoto and Swanson, 2016; Goswami et al., 2019; Kumar et al., 2016). OCT provides 3D surface images of different retinal layers. OCT-A provides 3D images of blood vessels using phase variance analysis (Migacz et al., 2019). The latter imaging modalities are non-contact and noninvasive thus can be used to collect time-dependent images for several imaging sessions for a long duration (Zhang et al., 2017).

Besides using noninvasive imaging techniques, temporal resolution between observations is essential factor for accurate fitting for the prediction of the tumor (Mehrara et al., 2009). Studies including more than 15 years of data are reported with much emphasis put on quantitative analysis of the outcome of prognosis, each with a not significant temporal resolution for individual cases (Cheng and Hsu, 2004). Inherently, human data cannot be expected to have a good temporal resolution because the patient needs to be treated immediately after diagnosis (Talkington and Durrett, 2015). The laboratory environment using animal disease models, however, facilitates the collection of time-dependent 3D images, provided that an accurate correlation can be obtained between human and animal imaging biomarkers (Gaupel et al., 2013; Goswami et al., 2019; Saxena and Christofori, 2013; Welsh, 2013).

Once the imaging is performed, post-processing tools are required to analyze the abnormalities. The character of the anomalies may vary case to case basis, thus requires a customizable soft tool. For example, the standard procedure of cancer imaging analysis involves delineation of tumor boundaries by experts (in case of human data and medical professional), segmentation of tumor bulge (in 3D), or neovascularization growth (if available) for qualitative estimation to design the radiation beam characters. It is termed as radiomics (Aerts et al., 2014; Parekh and Jacobs, 2019). As usual, the malignant tumor's phenotype, location with respect to the optical nerve, fovea, etc., and its depth inside the retina from choroid may vary according to disease. As eye cancer is rare, expertise to delineate the tumor is also not abundantly available. Animal models is one of the best alternate provide flexibility to create symptoms, study drug delivery mechanisms, test the efficacy of drugs, try new noninvasive imaging techniques, develop the soft tool, and train mathematical models to predict tumor growth, etc. The accuracy and time to delineate and segment certain abnormality vary from person to person and improves as more data sets are handled. This gain in expertise can be utilized by training the artificial neural network (ANN) based soft tool models. ANN and support vector machine (SVM) were shown to perform equally well if balanced learning and an optimized decision-making scheme ~~is~~ are being employed; otherwise, ANN provides better results (Ren, 2012). The performance of ANN, however, depends on the training data set and ~~employed~~ used model definition. It may perform inferior to SVM due to trapped into local minima solution or overfitting (Bisgin et al., 2018; Sakr et al., 2016). The performance of ANN and advanced deep learning-based techniques outperform SVM as the number of features becomes comparable to the number of samples (Ghorbani et al., 2016; Karaca et al., 2017). One can refer to the vast literature of limitation and utility of classical approach-based segmentation models (Lahmiri, 2017; Li Chen et al., 2009; Raju et al., 2018; Si et al., 2015; Taheri et al., 2010). Most of these works use human data for related analysis (Liu et al., 2019).

### *1.2. Motivation*

In this work, we are employing deep learning algorithms to identify the existence of tumor in OCT B-Scans and segment it automatically using mice eyes. Most of the works use human data lacking good temporal resolution (Litjens et al., 2017; Pekala et al., 2019; Wang et al., 2019). The use of the deep learning model for biomedical imaging segmentation is generally reported utilizing a particular/single model without justifying the rationale and analyzing suitability regarding diverse characteristics in data, number of images available for training/testing, etc. *This work highlights the need to carry out sensitivity analysis for available deep learning models prior to applying them to the data in hand.* Sensitivity analysis is performed with respect to types of tumors and the amount of available training data. Which index must be used to access the performance of models is also analyzed. It is possible that a performance index "A" indicates the betterment of a particular model M1, and at the same time, another index "B" declares another model M2 better. Thus sensitivity analysis regarding indices is also essential. Standard Performance index F-Score, Dice

Coefficient, Intersection over Union (I.O.U.), R.M.S.E., Loss, Converging speed are used as per according to common definitions (Munir et al., 2019). Another parameter, Hausdorff Distance (H.D.) defined as the maximum surface distance between the objects (in binary form), is also calculated.

## 2. Theory

Specialized deep learning architecture termed U-net and its existing variants (based on the type of encoders) are used in this work. Four models: (1) U-net with Vgg16 backbone termed here at UVgg, (2) hybrid of ResNet and U-net using Dice as a loss function (URsD), (3) hybrid of ResNet and U-net using Binary cross-entropy as a loss function (UResEn), and (4) hybrid model of U-net with Inception backbone (UIncp). We also played with its architecture by reducing the symantec gaps, but the performance was not considerable enough to include further complexity in the presented performance graphs (Long et al., 2014). Keras framework is used, which uses TensorFlow as its backend. We loaded the model with pre-trained weights, which were obtained by training the model on the ImageNet dataset. U-net (shape of network path evolves into U as down-sampling/encoder path forms the left-hand side of the U and the up-sampling/decoder path forms the right-hand part of the U) has shown the ability to efficiently extract the relevant features using relatively few training images (Ronneberger et al., 2015a, 2015b). The Vgg16 or (UVgg in this work) model has a rather large number of layers (X. Zhang et al., 2015). Many recent works have used the model with slight modifications. In one study, only 66 magnetic resonance (M.R.) images of patients were used for vessel segmentation resulting dice value of at least 0.76 (Livne et al., 2019). Retinal vessel segmentation was performed using public datasets (fundus images) DRIVE and STARE with greater than 95% accuracy using 40,000 patches of 48x48 image resolution extracted from 20 images of 565x584 pixels (Wang et al., 2019). 3D patches from the brain tumor segmentation (BraTS) 2018 challenge dataset (from 285 patients) were used to train a deep convolutional neural network (CNN) with a reported minimum dice score of 0.75 (Baid et al., 2020). Another variant of U-net, dense Inception U-net (with the addition of more layers), termed as UIncp, is used to segment a wide variety of multi-model noninvasive imaging data with at least 0.95 dice score (Cahall et al., 2019; Zhang et al., 2020). A residual neural network referred to as ResNet50 (made up of series of residual blocks) is also used. ResNet skips connections between two sequential layers realizing if the addition of the second layer is not improving the accuracy with overall flat gradients but slowing down the training process unnecessary (He et al., 2015). ResNet model resulted in 0.9 area under the curve (A.U.C.) values estimating Kirsten rat sarcoma viral oncogene homolog (K.R.A.S.) gene mutations status in patients suffering from colorectal cancer using 117 training C.T. images (He et al., 2020). The definition of the first four layers (used for downsampling) is from ResNet when hybridization of U-net and ResNet50 is obtained. During the final step of up-sampling, the model replaces transposed convolution process with shuffling the pixel. We termed this hybrid model as (a) URsD (if the Dice coefficient is used as loss function) and (b) URsEn (if binary cross entropy is used as loss function). Following are the definitions of standard performance indices with brief description:

*F-Score:* F-score is a weighted harmonic mean to estimate the relation between true positive ($TP$), false positive ($FP$), and false negative ($FN$) classified by the model. Eq. 1 describes the definition.

$$\text{F-score}=\frac{(1+\beta^2)(TP+\delta)}{(1+\beta^2)(TP+\delta)+\beta^2(FN)+FP+\delta} \quad (1)$$

Normalization factor $\beta$ helps to trade-off between precision and recalling factors. In this work, we have taken $\beta = 1$. F-score helps differentiate classifier performance if data has a class imbalance or unevenly distributed (Van Rijsbergen, 1979).

Dice Coefficient: Dice value gives similarity check between images from validation set (ground truth, $gt$) and predicted by deep learning model ($pr$). Eq. 2 describes the implemented definition:

$$\text{Dice coefficient} = \frac{2TP+\delta}{FN+FP+2TP+\delta} \quad (2)$$

I.O.U.: Intersection over union value is a metric that quantitatively penalizes single instances of bad classification as compared to F-score value differentiating worst-case performance by average performance of classifiers. The definition used in this work is given in Eq.3:

$$\text{IOU} = \frac{TP+\delta}{FN+FP+TP+\delta} \quad (3)$$

Where,

$$TP = \sum_{k=1}^{n} gt[k] * pr[k] \quad (4)$$

$$FP = \sum_{k=1}^{n} pr[k] - TP \quad (5)$$

$$FN = \sum_{k=1}^{n} gt[k] - TP \quad (6)$$

A slacking factor $\delta = 0.00001$ is used to avoid the singularity.

Hausdorff Distance (HD): is defined (Eq. 7) as

$$\mathrm{HD} = \max\{\mathrm{dhd}(\mathrm{gt},\mathrm{pr}), \mathrm{dhd}(\mathrm{pr},\mathrm{gt})\} \quad (7)$$

where,
Directional Hausdorff Distance ($dhd(\mathrm{gt},\mathrm{pr})$) is defined in Eq. 8:

$$\mathrm{dhd} = \max \mathrm{gt}_0 \text{ c } \mathrm{gt} \left[ \min \mathrm{pr}_0 \text{ c } \mathrm{pr} \left[ ||\mathrm{gt}_0 - \mathrm{pr}_0|| \right] \right] \quad (8)$$

The $dhd$ intuitively finds the point $gt_0$ from the set $gt$ that is farthest from any point by $pr$ and measures the distance from $gt_0$ to its nearest neighbor in $pr$.

# 3. Materials and Methods

## 3.1 Deep Learning Model implementation

The flow diagram of this work is shown in Figure 1. Adam optimizer is used. Hyperparameters such as learning rate (L.R.), epochs, and classes (1 class only, i.e., tumor) are used. It is found that the performance parameter converges well before 100 epochs are over. H.P. Z4 workstation having Intel® i9 7840x processor with 64 GB RAM, Nvidia Quadro P4000 8 GB GDDR5 GPU is used to execute the codes (written in Python). The codes with a readme file, trained weights, and sample data are attached with this paper.

We have used MedPy, a python library to deal with medical image processing, to extract both the plain images and their corresponding mask (tumor) images from the OCT scan files.

This library provides access to several functions for preprocessing. A brief description is given below:

1. Conv2images and MedPy are used to manage the data set,
2. Conv2bw is used to convert the greyscale mask to black and white so that either the pixel will be 255 (white) or 0 (black). This is necessary because our implementation of the model then normalizes the whole image by dividing it by 255. It converts pixel values 255/255 = 1 (representing tumor) and 0/255 = 0 (representing no tumor).
3. Give-counter is used to estimate how many pixels have value > 0. It helps not to include completely black images after certain numbers in the training process. Black images may steer the model to learn nothing. So we kept the threshold to 50.
4. Preprocess function resizes all the images to a certain dimension (x0.25 in our case while maintaining the OCT system aspect ratio in 3D). It also converts data into four channels as the deep learning model requires 4 image channels (batch size, red, green, blue). Image and batch size (8 in our case) depend on data processing system / P.C. configuration.
5. Handling generators are used to optimize the heavy data that load files which are required

by training functions at a particular time. It also helps vectorization of data. We here applied some data augmentation techniques (passed as a parameter, as a dictionary format). We did not apply any augmentation on the validation set.

The images were then partitioned into three groups which we referred as Training, Testing, and Validation set. The model was trained on the training set and was validated using the validation set, we didn't touch the testing set to tune our hyper-parameters, and it was only used once we were satisfied with the model.

The primary section of each model architecture is responsible for extracting the characteristics/details from the images. This part consists of a stack of Convolutional, Max Pooling, and Dropout layers. As we go down the path, the height and width of the volume decreases, whereas the depth increases (no of channels). The initial convolution layer contains filters having a very small receptive field: 3×3 meant to capture the notion of left/right, up/down, center. 1x1 convolution layer can be seen as a linear transformation of the input channels. After each filtering step, in the convolution loop, padding is performed to preserve the spatial resolution. The depth of convolution layers makes each architecture different. The number of backend layers and the number of trainable parameters in Vgg16, Inceptionv3, and Resnet are 66, 352, and 231 and 23, 748, 241, 29, 896, 689 and 32, 513, 556, respectively.

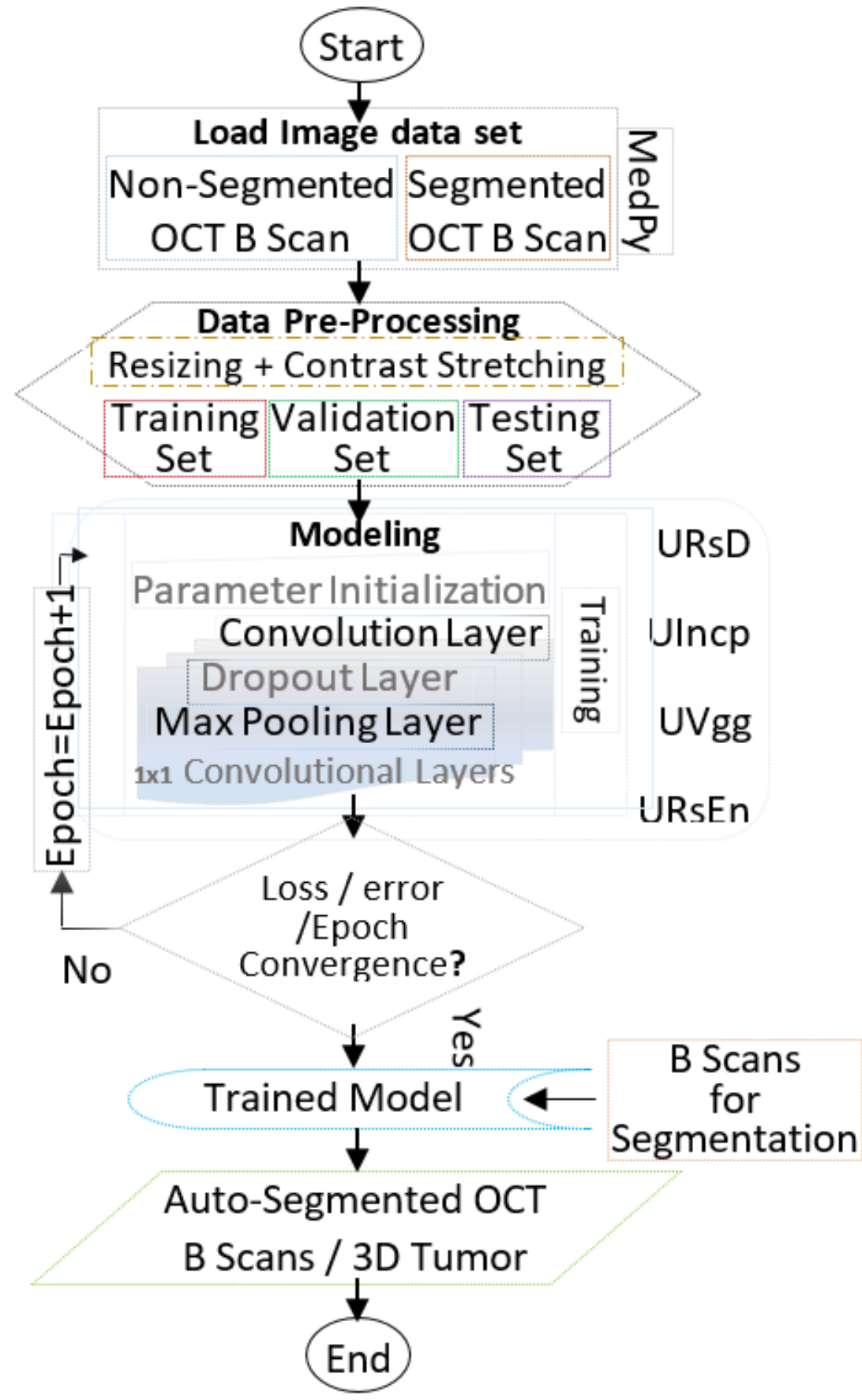

Fig. 1. Flow diagram

The second part is an expansion part (or decoder). This portion uses the transposed convolutions with skip connections coming from the encoder. It helps to get the spatial information that might have got lost during the encoding process. The decoder then provides us with the segmentation mask. It is then compared with the true mask, and the loss is then calculated, which is back-propagated to update weight so that we can get a better mask. This procedure is done till convergence is achieved. In calculation step, we took an average of performance indices values over batch. Exact values were used without rounding off the predictions.

### 3.2 Data Preparation

We have borrowed data (tumor data) from previously published work (Goswami et al., 2019) and from our own laboratory (healthy mice eyes). Hybrid compact OCT system (132 nm band centered on 860 nm) collects A-scans (single tomographic axial samples) at 100 kHz, with an axial resolution of ~2 µm having x-, y-analogue resolutions of ~3.5 µm. A standard OCT volume sampled a retinal area of 1.6 mm × 1.6 mm (assuming an angular conversion factor of 32 µm/deg.) and consist of 796 B-scans (horizontal scans), with each B-scan, in turn, comprising 2000 A-scans (axial backscatter profiles). Six spatially consecutive B-scans were averaged to reduce the speckle noise in the OCT images (P. Zhang et al., 2015).

The tumor data (50 mice eyes) contains tumor progression follow-up from baseline/control until the animal was euthanized under ethical committee guidelines. Glioblastoma xenograft was used considering its similarity with the symptoms of affecting the central nervous system (C.N.S.) after the occurrence of melanoma (Arcega et al., 2015; Scarbrough et al., 2014). The growth of tumor volume was categorized as exponentially growing, logarithmically saturating, and following a Gaussian distribution with respect to progression of days after xenograft. In published work, data were segmented using a support vector machine (SVM) based semi-automatic segmentation tool (Fuller et al., 2007). It is found that segmentation had minor errors; minimizing those would take significant time. Typical time and accuracy for segmentation using such a tool vary according to (a) experience of the user, (b) size of the tumor, (c) its morphology (flat or bulged, fragmented or continuous, big or small tumor), (d) data suffering from incidences when the mouse was breathing significantly, (e) tumor location near or far from the choroid, and (f) thicker blood vessels in N.F.L. region causing shadowing effects, etc. Several such characters are shown in Figure 2. The scanning scheme is illustrated in Figure 2(a). The axial collection of A-scans generate a single B-Scan or lateral collection generates *en-face*. Stacks of *en-face* or B-scan give 3D image of inner structure where light was focused. Figure 2(b) shows OCT B-Scan of a healthy retina (in grey color). It is merged with respective OCT-A B-Scan (in red tint) in same image, depicting existence of motion of bodily particles causing shift in phase. The Figure 2(c) taken right after xenograft. It shows the injured retina above to retinal pigment epithelium (RPE) which sometimes leaves its mark even after retina gets healed naturally. Figure 2(d) shows a retinal detachment. Figures 2(e) and 2(f) show small and very big tumor (regions highlighted by yellow color with red tint of OCT-A in background).

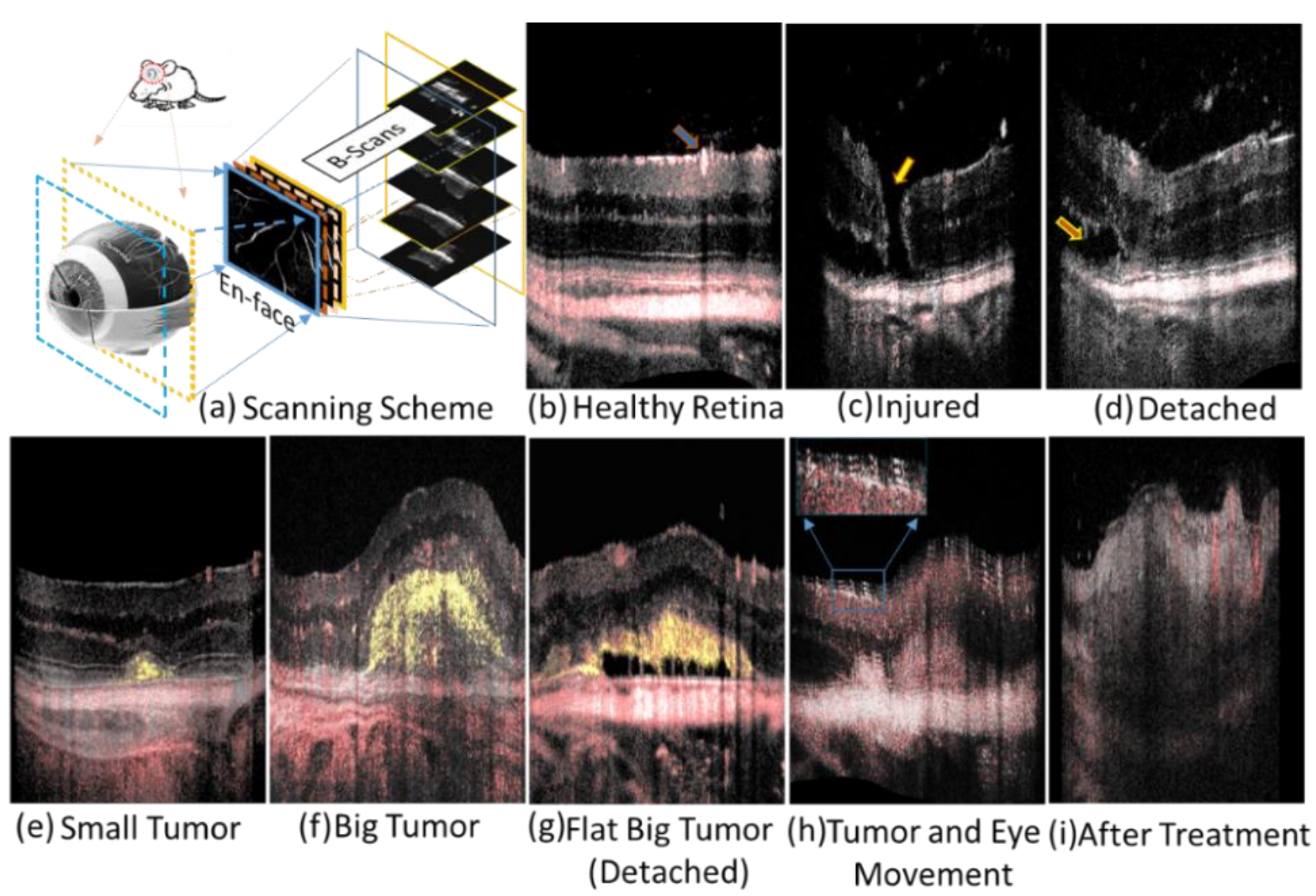

Fig. 2. Types of abnormalities in OCT Data set after Xenograft

Similarly, a big flat tumor is shown when it is detached from RPE from the central region. It is, though, attached with RPE in several fragmented areas. Figure 2(h) shows the effect of mouse breathing on image quality. The mouse eyeball remains stationary under anesthesia. It, however, develops slight resistance against the previously used threshold flow rate. This behavior affects the ability to keep the eyeball immobile after multiple imaging sessions. The movement gets reflected into B-scans representing a single retinal layer into numerous. OCT images also show significant speckle noise. There are standard image registration, temporal speckle-averaging steps to minimize (but not to zero) these effects (Zhang et al., 2019). Figure 2 (i) shows a big tumor after several treatment sessions. In this image, it is shown that due to treatment retinal layer structure gets all meshed up. Figures 2(b) and 2(g) also serves the purpose to highlight shadowing effect of one of the major vessel in Nerve Fiber Layer (N.F.L.). All images in Figure 2 are not from the same animal, each having different natural eye curvature. The flattening step is performed for uniform comparison. We chose not to follow the flattening step in segmentation analysis to maintain the natural curvature of the eyes. For better visual understanding, in 3D, one can refer to Movie A(Goswami, 2020a).

The entire data were annotated manually using the ImageJ tool (Rueden et al., 2017). Duration for tumor manual segmentation is recorded for 4 different people (2 females and 2 males). The duration ranged between 35 minutes to 1.20 hour per eye. All four people used computers of same configuration. Each retinal OCT scan contains 359 images of 540x1026 pixels. The duration improved between 20 – 50 minutes per eye after the same person segments at least 20 eyes. Delineation and segmentation is certainly a very time taking process. OCT-A scans cannot be segmented manually. The edge detection process can be performed on manually segmented OCT intensity data. The extracted edge index thus can help to segment OCT-A data. The efficiency of edge detection depends on speckle noise; however, it affects the overall segmentation of OCT-A data.

## 4. Results

Figure 3 presents the primary segmentation steps. Figures 3(a) – 3(d) illustrate the stack of OCT B-Scan in the 3D form in grey and colorized scale. One of the extracted B-Scan in Fig. 3(a) shows a small tumor with minor detachment penetrating RPE and remaining well below the outer plexiform layer (O.P.L.), affecting the outer nuclear layer (O.N.L.) and ellipsoid zone (IS / OS junction). Figs. 3(b) and 3(c) show its extraction from the partial section of the 3D stack in grey (usual depiction in literature) and colorized version (adds depth perception and surface roughness). Figure 3(d) shows a full 3D retinal structure. Tumor with respect to optical nerve location at origin/reference point from the 3D axis (blue color) is shown. The top part from N.F.L. till tumor surface is made translucent for better visualization. The color scheme is arbitrary. Bright red color with a slight yellowish tint is added to depict the surface variation of the tumor. The pale orange color refers to the structure of RPE. Figures 3(e) to 3(j) show segmented tumor region by hand (True), SVM, and Deep learning models. Grey color scale for OCT and red tint is used for OCT-A data. It is shown that all 4 segmentation methods include RPE as part of the tumor. It highlights the biggest challenge for the deep learning algorithm. That is to segment tumor boundaries when it's getting merged/fused with RPE. The presence of abnormalities shown in Figure 2 affects the performance. Besides this, SVM includes small disconnected dots (between N.F.L. and the top surface of the continuous bulge of tumor) as part of the tumor. These dots are marked by arrows in the same figure. Auto-segmentation shows presence of discontinuous elongation of the tail of the tumor (highlighted by arrow).

This particular B-Scan segment created by UIncp (fig. 3(i)) seems to be most similar to the True case. Figures 3(k) - 3(p) include 3D profiles (stacked B-Scans) of tumor only. In these figures, all other parts of the retinal structure are suppressed. The augmentation of erroneous extension of tail and fragments/dots from each B-Scan added up to a faulty tumor structure in the deep learning model except in the case of URsD. SVM seems to mimic the true 3D structure of the tumor with a minor presence of dots.

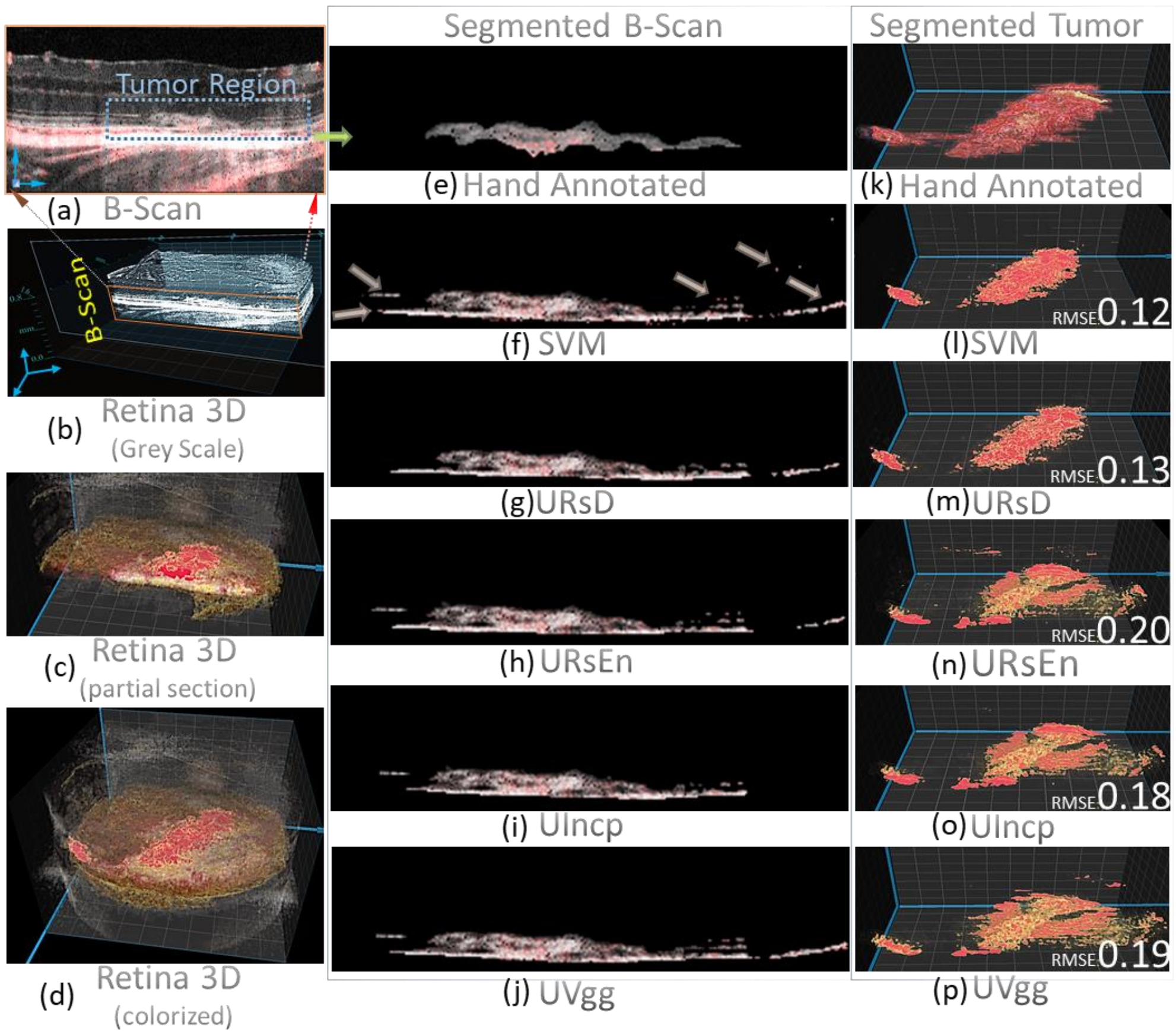

Figure 3: Mouse retina visualization and segmentation strategy using various models

Respective R.M.S.E. values are shown in the same figure for quantitative comparison. Visually, URsEn seems to be giving the most faulty segmentation. However, the difference between R.M.S.E. between URsEn and URsD is only 0.07, justifying the need to include other performance indices for sensitivity analysis. Only 40 hand-annotated images are used to train the deep learning models.

### 4.1. Sensitivity Analysis

An exhaustive study is carried out to identify the best model suitable for tumor segmentation now. We chose to test the model's capability with respect to the number of images (with sparse variation) used and several different data sets (with low variation: benign tumor and high variation: in case of treated tumor images) used during the training and testing phase. We note that as per according to the definition of Dice coeff., F-score, and I.O.U. values must be high and approach to 1 as the number of training images (N-Train) increases. For reliable analysis, their values again remain high no matter how many number of instances/cases/testing images (N-Test) are tested. The R.M.S.E., loss coefficient values must remain low and decrease. Figure 4 shows performance parameter plots for the first category of data (similar size and shape of tumor) using all four deep learning models. It is achieved by segregating B-Scans of untreated eyes from the same mouse. However, abnormalities shown in Figs. 2(b)-2(h) are present naturally. The minimum Dice coefficient value is estimated 0.73 when 462 training images and 66 testing images are used by the URsEn model.

Dice values for the UIncp model are relatively greater than other models. The maximum Dice coefficient value is estimated 0.85 when 3230 training images and 461 testing images are used by the hybrid URsD model. Better classifier performance is shown by URsEn as far as F Scores and I.O.U. are concerned. UIncp shows the next best results. Best loss values (near to -1) are given by URsD showing its compatibility with the objective function. However, loss values by URsEn also remain flat and near to desired value (0 in case of cross-entropy). R.M.S.E. values by URsEn and UIncp remain lowest. R.M.S.E. and H.D. plots show a diverging pattern by UVgg. It also shows good loss performance up to a certain threshold beyond which UIncp takes over. H.D. values seem to be going up by UIncp as well. It is evident from all of the plots that quantitatively, the performance of UVgg is inferior to other models.

We have linearly fit each plot according to Eq. 9 to analyze which plot would have desired decay when larger (than used currently) data sets would be used. For sharp linear fit $P_{10}$ must be high and $P_{01}$ must be low.

$$f(\text{N. Training}, \text{N. Testing}) = P_{01}\text{N. Testing} + P_{10}\text{N. Training} + P_{00} \quad (9)$$

The values are shown in Table 1. Desired values are highlighted (red best and orange second best). Performance of UIncp is found saturating with large data set (better visible in 2D plots in shadow region). This analysis indicates that plots by URsD have a better slope, and it may outperform any other model if a large number of data sets are used. The Model URsEn also has a comparable and desired slope. Overall, for this range and data set combination (without treatment and less characteristic variation), URsEn (cross-entropy as Loss function) is the best predictive model as far as similarity with hand-annotated data is a concern. The next alternative is UIncp.

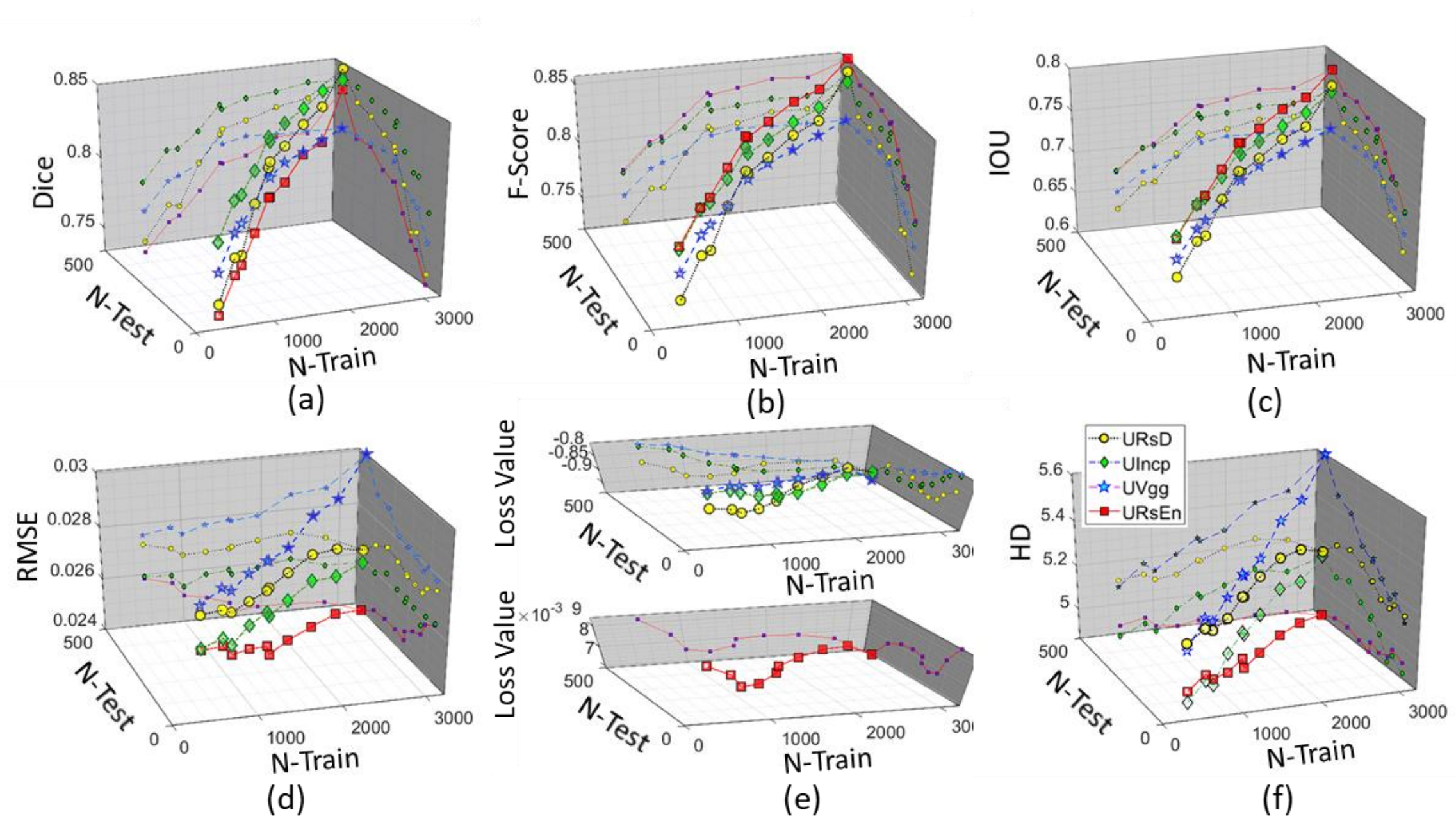

Fig. 4. Sensitivity analysis with data from category 1: mean values of performance indices are plotted with varying values of number of images in training (N-Train) and testing (N-Test) set.

Table 1. Coefficient of performance indices fitting for data with treatment and less variation

| Model | URsD | | UIncp | | UVgg | | URsEn | |
|---|---|---|---|---|---|---|---|---|
| Coefficient | $P_{10}$ | $P_{01}$ | $P_{10}$ | $P_{01}$ | $P_{10}$ | $P_{01}$ | $P_{10}$ | $P_{01}$ |
| Dice | 0.0042 | -0.0294 | 0.0028 | -0.0195 | 0.0031 | -0.0215 | 0.0028 | -0.0196 |
| Loss | -0.002 | 0.0119 | -0.0056 | 0.0390 | 0.0002 | -0.0019 | -0.0019 | 0.0001 |
| RMSE | 0.0009 | -0.0062 | 0.0007 | -0.0050 | 0.0007 | -0.0051 | 0.0007 | -0.0055 |
| F-Score | 0.0046 | -0.0321 | 0.0032 | -0.0224 | 0.0034 | -0.0237 | 0.0037 | -0.0261 |
| IOU | 0.0048 | -0.0336 | 0.0035 | -0.0246 | 0.0038 | -0.0262 | 0.0042 | -0.0295 |
| HD | 0.0062 | -0.0431 | 0.0101 | -0.0702 | 0.0016 | -0.0100 | -0.0020 | 0.0140 |

In the next stage, a similar analysis is performed, using relatively large variation in the data set, a mix of all kinds of tumors and characters, including eyes, have gone under treatment protocol. Plots containing performance analysis from all models are shown in Figure 5. Speed analysis of models achieving convergence in loss function with respect to epochs ($ep$) is also performed. We found all of the performance parameters vs. epoch plots a good fit to an exponential function, Eq. 10.

$$(p = ae^{-ESR\times(ep)} + c) \tag{10}$$

In this function, $ESR$ termed as saturation rate, indicating the rate by which a parameter $p$ achieves steady-state value. Figure 5 shows that for all models (except URsEn), performance indices have minor differences as compared to the previous case when data has no variations. However, UVgg is good if the training data set has less than 1519 images. URsEn shows relatively poor performance as far as F-Score, I.O.U., and R.M.S.E. are considered when less than 1519 images are used in training. It has a relatively higher saturation rate value, indicating faster convergence and outperforming other models afterward. Dice values of URsEn always remain relatively poor than any other model.

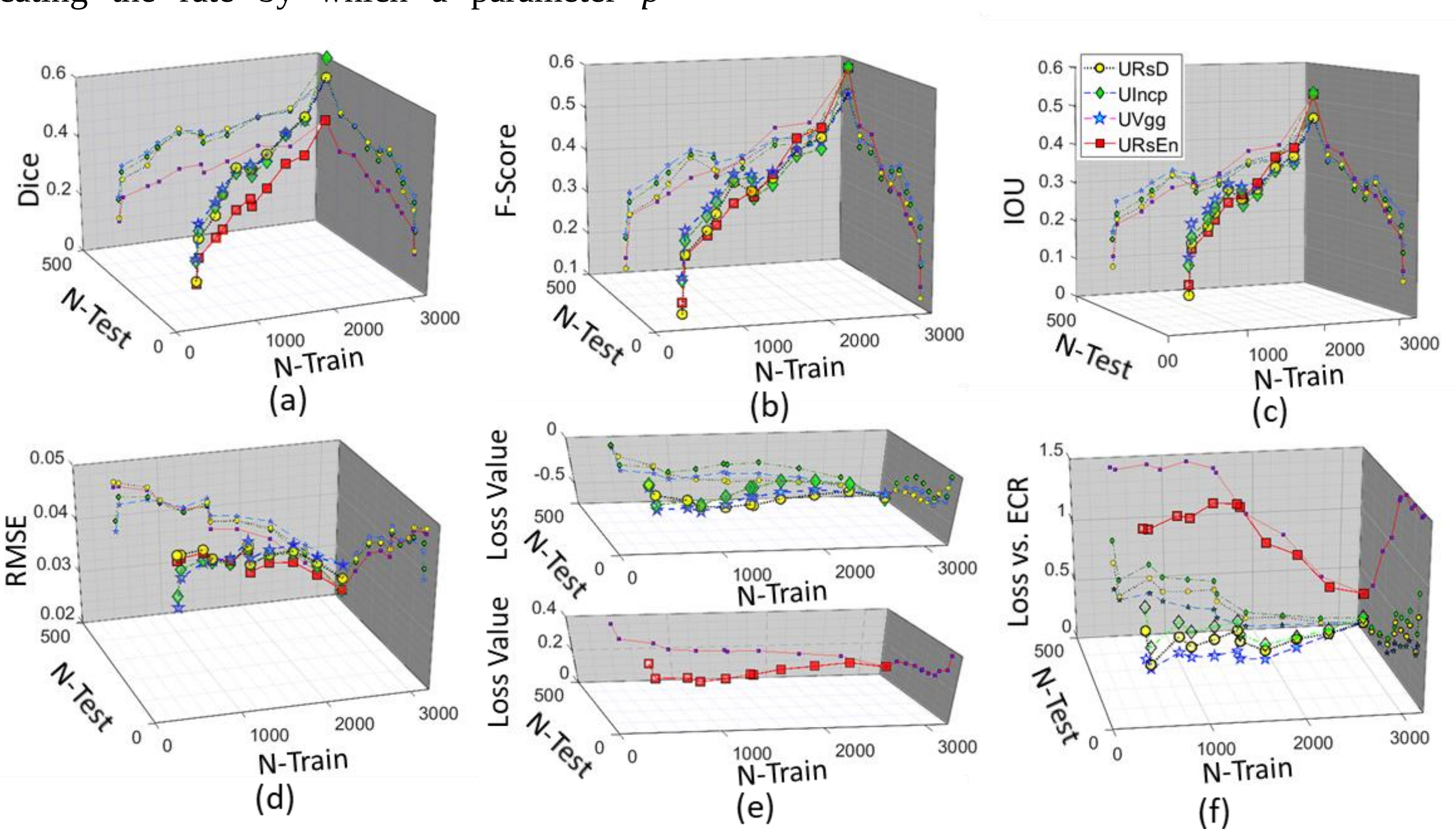

Fig. 5. Sensitivity analysis with data from category 2: various mean values of performance indices are plotted with varying values of number of images in the training (N-Train) and testing (N-Test) set.

Table 2. Coefficient for performance indices fitting model using variation analysis data

| Model | URsD | | UIncp | | UVgg | | URsEn | |
|---|---|---|---|---|---|---|---|---|
| Coefficient | $P_{10}$ | $P_{01}$ | $P_{10}$ | $P_{01}$ | $P_{10}$ | $P_{01}$ | $P_{10}$ | $P_{01}$ |
| Dice | -0.0048 | 0.0340 | -0.0057 | 0.0409 | -0.0047 | 0.0337 | -0.0019 | 0.0141 |
| Loss | -0.0010 | 0.0056 | 0.0247 | -0.1742 | 0.0205 | -0.1445 | 0.0037 | -0.0267 |
| RMSE | 0.0009 | -0.0062 | 0.0007 | -0.0050 | 0.0007 | -0.0051 | 0.0007 | -0.0055 |
| F-Score | -0.0052 | 0.0369 | -0.0055 | 0.0391 | -0.0043 | 0.0307 | -0.0044 | 0.0320 |
| IOU | -0.0059 | 0.0423 | -0.0053 | 0.0376 | -0.0046 | 0.0329 | -0.0038 | 0.0277 |
| HD | 0.0533 | -0.3807 | 0.0347 | -0.2492 | 0.0516 | -0.3649 | 0.0458 | -0.3288 |
| ESR | 0.0164 | -0.1165 | 0.0136 | -0.0973 | -0.0037 | 0.0251 | 0.0178 | -0.1280 |

## 4.2 Reproducibility Analysis

We ran the training codes using the same data set multiple times (8 re-runs/trails), testing which model would experience the “reproducibility crisis”(Hutson, 2018; Raff, 2019). Dice, F-Score, R.M.S.E. and Loss values are plotted (shown in Figures 6(a)-6(e)), using data set in category 1 and for all four models, URsD, UIncp, UVgg, and URsEn in clockwise orders with red, green, pale yellow/beige and blue color of boxes. Table 3 shows respective index values that are used in the x-axis (the notation is used to achieve compactness and clarity). Figures 6(a)-6(c) show that median Dice, F-Score, and I.O.U. values improve (as desired) for all models. The box size and length of whiskers reduce with size in the training data set. Plots for URsEn show a relatively bigger size of boxes, indicating that it suffers most from reproducibility issues. It is shown that with the number of images (testing and training, both) the reproducibility in output improved if Dice, F-Score, and I.O.U. values are considered. This conformity in results, however, deteriorated if R.M.S.E. and loss values are considered. A bar plot also shows mean values of standard deviation in Figure 6(f). It is inferred that UIncp and URsD are the most stable models. It is also shown that loss is the most sensitive and R.M.S.E. is the least sensitive performance indices (reconfirming visual observations obtained in Figure 3) to carry out deep learning model testing. Dice, F-Score and I.O.U. exhibit comparable stability.

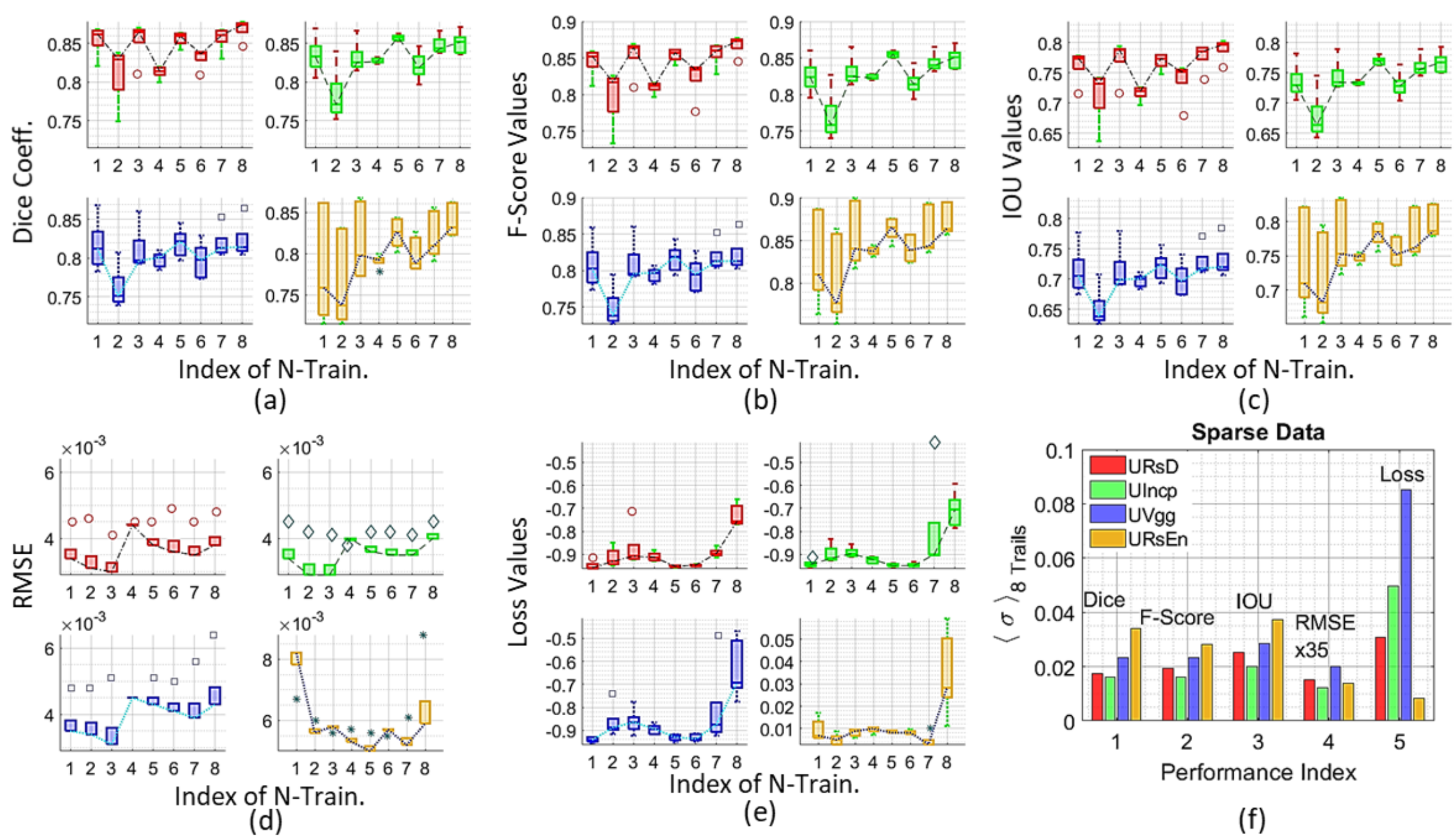

Fig. 6. Sparse data and reproducibility analysis: each model is trained and tested 8 times using the same data with fewer variations. Figs. 6(a)-6(e) show maximum, minimum, 1st, and 3rd quartile (top and bottom

edges of box) and median lines. Fig. 6(f) shows the mean values of standard deviation for five performance indices. The color scheme for the bar chart and plots are the same.

Table 3. Index of N-Train and actual values in Fig. 6 and Fig. 7

| Index of N-Train | 1 | 2 | 3 | 4 | 5 | 6 | 7 | 8 |
|---|---|---|---|---|---|---|---|---|
| Actual values of N-Train | 402 | 801 | 1229 | 1519 | 1879 | 2296 | 2739 | 3230 |

A similar analysis is presented in Figure 7 using data defined as category 2, i.e., with treated eyes. It is shown that the reproducibility crisis gets resolved significantly for URsEn relatively than other models as it is observed for category 1 data analysis. In fact, it becomes better beyond a certain number of images (1519) used for training. Overall performance by UVgg exceeds other models as far as reproducibility is concerned. Relative sensitivity of R.M.S.E. to access the performance of model improved for data variation analysis as compared to category 1 data analysis. However, it remains inferior to other indices. The loss remains most sensitive of all, again.

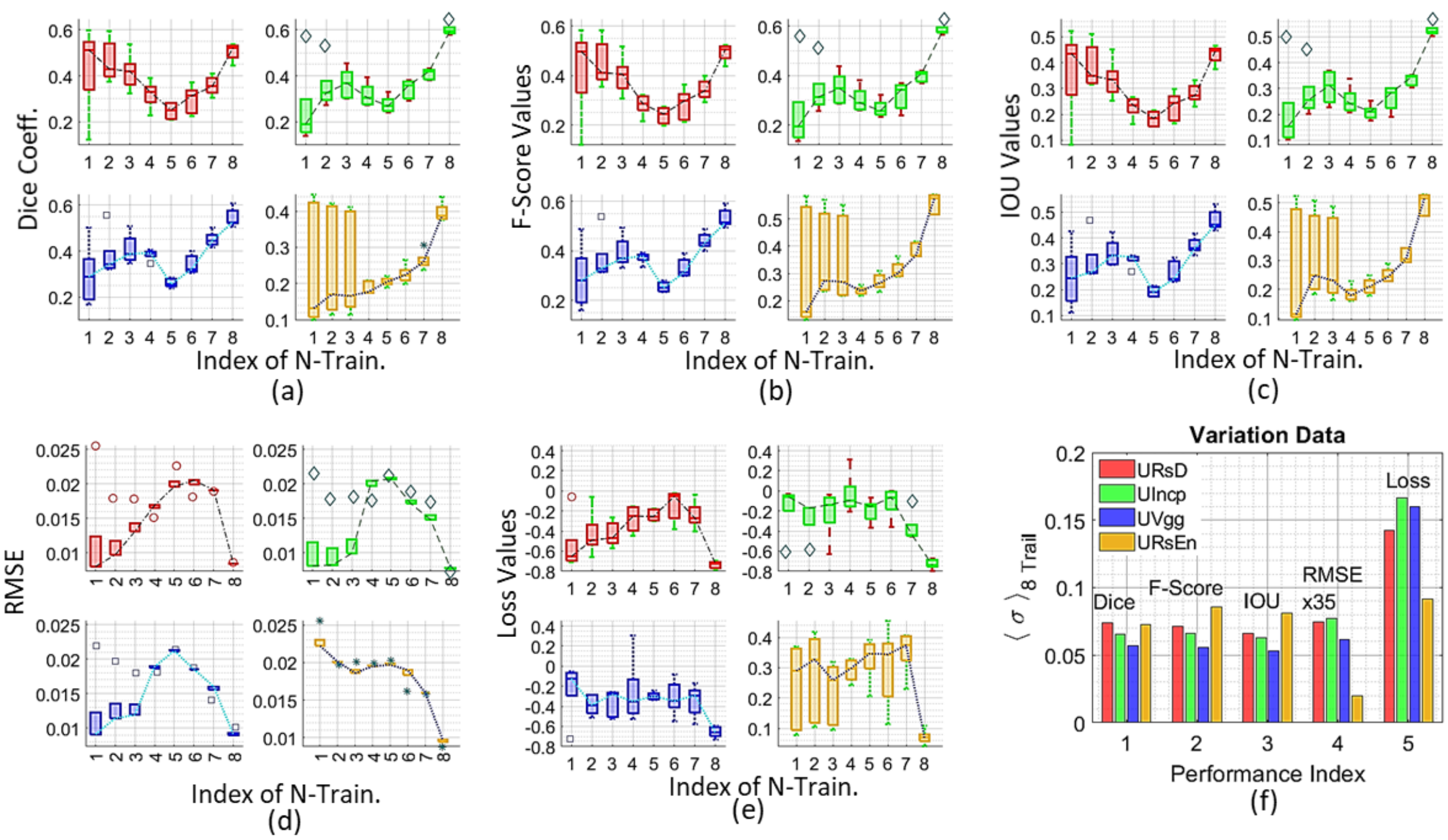

Fig. 7. Data with variation and reproducibility analysis: each model is trained and tested eight times using the same data with variation in characters. Figs. 7(a)-7(e) show maximum, minimum, 1st, and 3rd quartile (top and bottom edges of box) and median lines. Fig. 7(f) shows the mean values of standard deviation for five performance indices. The color scheme for the bar chart and plots are the same.

### 4.3 Visual conformity analysis

Figure 8 presents an example of one case of tumor which is imaged right after imaging assisted photodynamic treatment process. The detail of this specific data is given in next section. Figure 8(a) shows hand-annotated segmented 3D volume as well as four B-scans (extracted at a regular distance from each other) in the first column. Columns 2, 3, and 4 depict automatically segmented images using UIncp, URsD, UVgg, and URsEn, respectively. The first and second rows (respective B-scans are tagged by 1 and 2) contain relatively big part of tumor near to optical nerve. Hand annotated segmentation matches with segmented images from all methods for the first row. However, deep learning methods fail to separate RPE from the tumor. We note that the part of the retina shown in B-scans in the first row was not exposed by the LASER beam. Second row contains a part that fell under LASER's focus blasting the outer shell, thus releasing Nano-Dox, locally. It created a mashed retina (major difference between treated and untreated retinal layers, which is also depicted earlier in Fig. 2(h)). Visual inspection indicates that segmented images by UVgg and URsEn are relatively similar to the hand-annotated image than segmented by UIncp and URsD. Row 3 contains a very thin layer of tumor and segmentation by URsEn, and UIncp matches only. Hand annotated B-Scan images in the final row show absence of a tumor. Except for URsEn, every deep learning method is showing a slight presence of the tumor. Close observation reveals that for the last row, it is possible that the image does have a presence of a very thin tumor or damaged retina. Here, UIncp appears to outperform hand-annotated results. Model URsEn is closest to the hand-annotated image.

We conclude that UIncp would be the best universal model for this particular data and tumor segmentation problem, irrespective of variation in data or its size. URsEn is the second best option, provided that the reproducibility crisis is neglected.

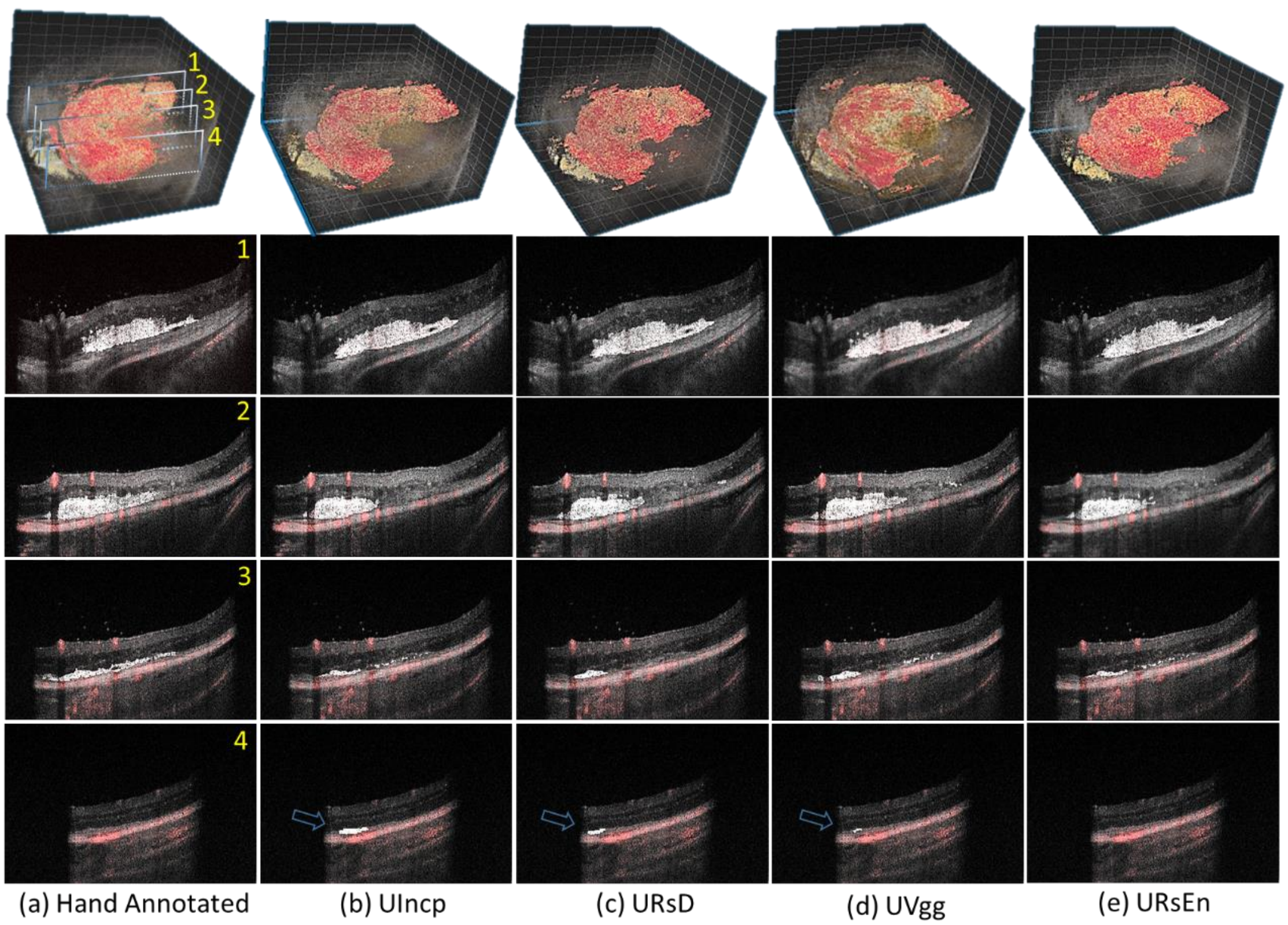

Figure 8. Performance comparison of models for treated tumor: column 1 has a hand-annotated B-Scan, rest of the columns show respective segmentation by Deep Learning models, segmented tumor boundaries (white color) in hand-annotated images matches by auto segmentation performed by URsEn (last column).

If a larger amount of data is available then URsD is shown to be performing better. UVgg can be preferred if data contains tumors after treatment. Finally, it is used in the next section to segment time series OCT tumor volume to study the effect of a cancer treatment protocol.

### 4.2. Tumor Growth Estimation

Best candidate data (in terms of variations) from an athymic nude - Foxn1nu mouse (injected with 250 glioblastoma cells dissolved in 0.25 microliters) undergone for 7 sessions of treatment is used to exhibit the application of segmentation ("Athymic nude mice: Hsd: Athymic Nude-Foxn1nu mutant mice," n.d.). Figure 9 contains OCT B-Scans, 3D (perspective view), and top view of tumor surface growing over the period of 129 days (considerable temporal resolution) of follow-up after xenograft. The deep learning model used for this analysis contains 7975 images (with 3230 images containing tumor) for training, 461 images for testing, and 922 images for validation.

We note this is the first time tumor growth with long temporal resolution using in-vivo, non-invasively modality is shown. The color scheme is as per according to the reflection of light. On greyscale, if the tumor is dense or its structure is meshed up / retinal layers are not clear, it is visible as grey, and we showed it using red color. The blue color is used to show if the tumor is translucent had an absence of matter. Overall, it adds to the depth perception. On greyscale, it was visible light grey. The yellow color is used for the middle range. This color scheme is chosen/fixed such that the overall result shows the tumor surface with better visualization. The red color is used for the tumor surface. Figure 9(a) shows that on $7^{th}$ day, the shape of the tumor is flat (barely touching the O.P.L., only affecting O.N.L. and penetrating RPE) and is made of two parts: one at the central region and another at the corner part of the view. The B-Scan does not show the corner part indicating the difference in depth. The shape of the tumor is shown to be transformed from flat to round and relatively bigger on the $33^{th}$ day (shown in figure 9(b)). Figure 9(c) shows that the tumor is proliferating towards the optical nerve with a slightly detached region, forcing it to penetrate O.P.L. The first treatment session (protocol: imaging assisted photodynamic nanodrug (via tail-vain injection)) was performed after 93 days from xenograft and every third day afterward till $118^{th}$ day. Figures 9(d)-9(g) show OCT 3D images of the tumor right after the treatment session. The tumor seems to gain its mass, and shown in Figure 9(e).

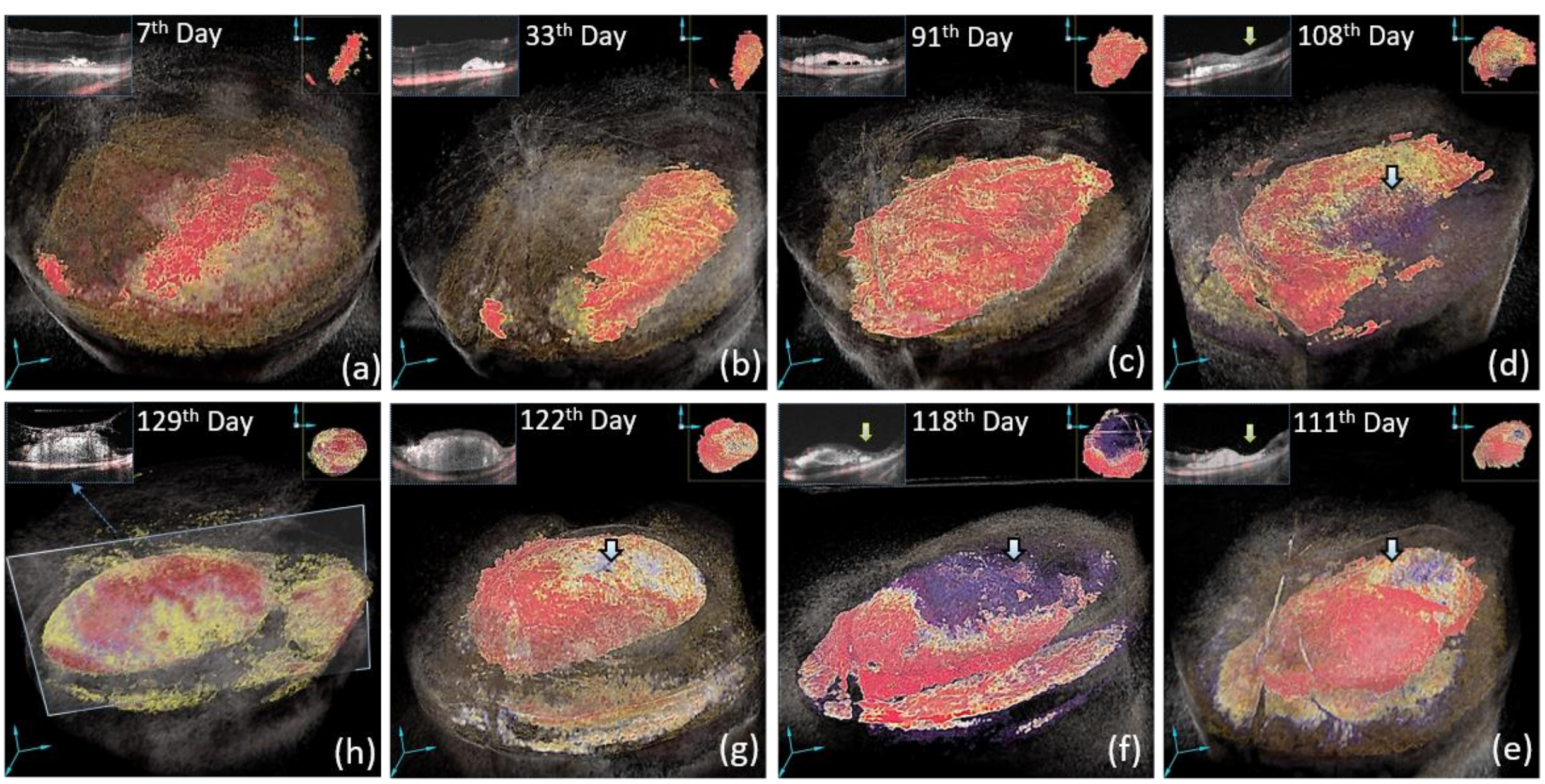

Fig. 9. Tumor progression: time scale growth of the tumor is shown from the $7^{th}$ day after xenograft (a) till (e), treatment starts after $93^{rd}$ day. Blue color represents the absence of matter, red and yellow are used for depicting gradient/roughness of the surface. Inset top left is one of the B-Scan (the corresponding location is shown in Fig. 9(h)), and the inset top right shows a top view (X-Y).

The LASER beam pointer is shown in all of the treatment figures by an arrow heading downwards. The final image shown in Fig. 9(h) was taken when it was decided to euthanize the mouse as the tumor was about to touch the lens (also visible in the image). Figure 9(h) shows the plane from which B-Scan is extracted. Full data is shown in the form of Movie B(Goswami, 2020b). The above qualitative analysis depicts the inefficiency of treatment protocol showing that after treatment, the tissues bulge in the tumor instead of dying outgrown. In the histopathological analysis, we found this was not true. To clarify, the normalized segmented tumor volume obtained by OCT Angiography and OCT intensity modality is estimated and shown in Figure 10. The growth of OCT-A follows the growth estimate OCT except at few points. OCT-A volume is higher than volume by OCT between 18$^{th}$ till 51$^{th}$ day and 81$^{st}$ till 93$^{th}$ day. OCT-A volume reflects growth in a blood vessel inside this tumor. Treatment was started after the 93$^{rd}$ day. It is observed that after the 51$^{st}$ day, growth in neovascularization driving overall growth in tumor volume (tracked by OCT). This logic gets further strengthened by another observation. Once the treatment started after the 94$^{th}$ day, the OCT-A volume decreases sharply, affecting the tumor volume.

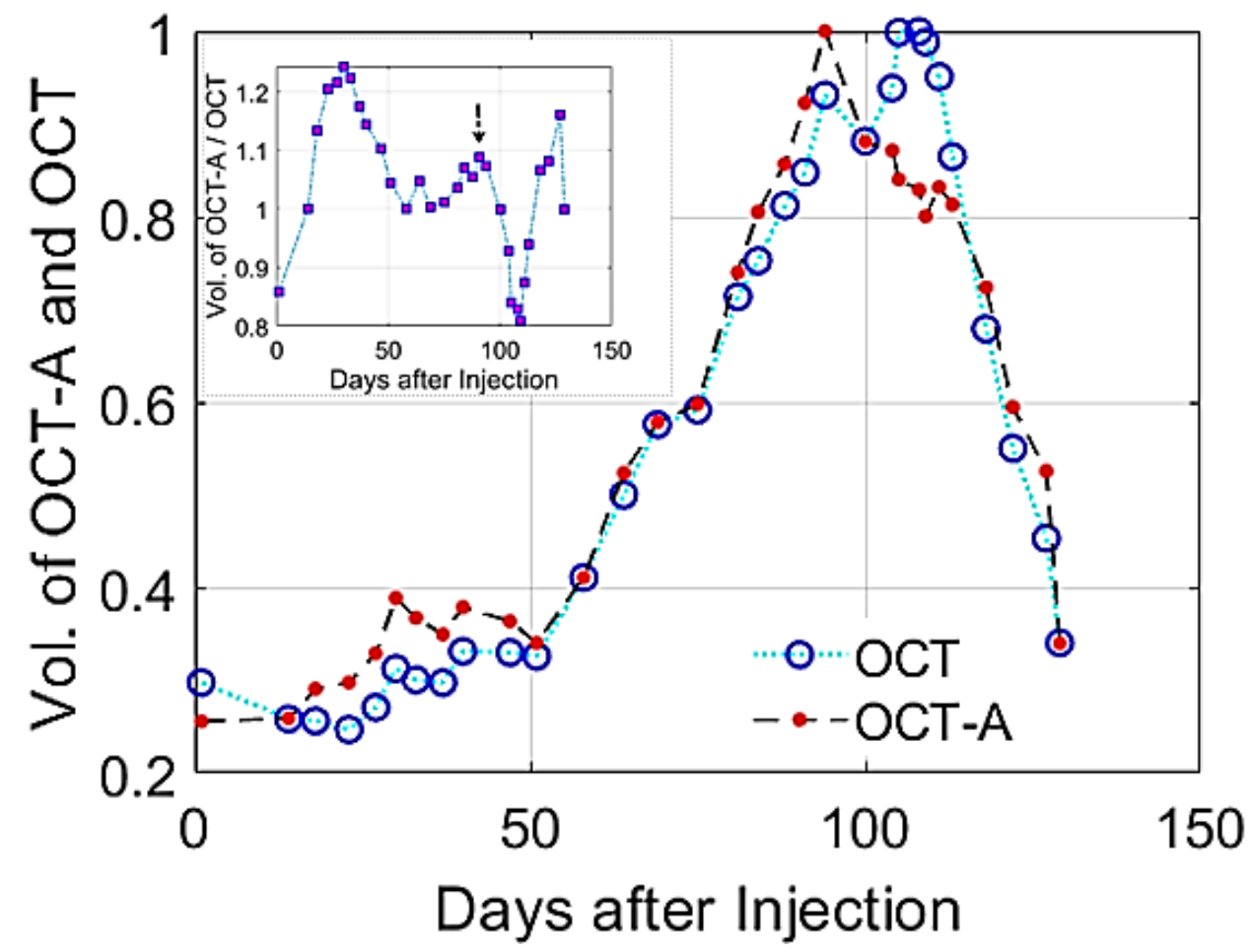

Fig. 10. Volume Growth Estimation of surface and blood vessels

This statement is supported by a ratio in between volumes by OCT-A and OCT. This ratio is plotted in the inset in Figure 10. The plot in Fig. 10 shows that tumor volume indeed decreases as the protocol progresses. However, Fig. 9 shows tumor expansion. We note that Fig. 9 depicts the outer surface of the tumor in 3D. It is mostly hollow from the inside. This is also shown by B-scans and a blue color scheme. The treatment protocol gives a controlled Gaussian profile to the growth of tumor volume. The volume is calculated by normalized values of number of pixels in the segmented tumor region.

## 5. Conclusion & Discussion

Auto-segmentation schemes using deep learning models are discussed. OCT Data for artificially induced cancer growth is studied. It is shown that Nano-Dox drug treatment helps to control the growth of tumors, but Photodynamic imaging assisted treatment therapy transforms tumors into cyst-like structures. Dice, F-Score and I.O.U. as indices show similar sensitivity to access the performance of deep learning models. Loss value is the most, and R.M.S.E. is the least sensitive performance indices to access the sensitivity of deep learning models. U-net with binary cross-entropy as loss objective (URsEn) is the fastest to converge, model, especially if one has no access to GPU for training the model.

It is shown that URsEn performs as par to other models for data with less variation/class imbalance; however, it suffers reproducibility crisis. It also has a drawback that its performance remains sub-optimal than others up to a certain threshold value of training and testing data set if data has large variation/class imbalance. Finding this threshold value will add to the processing time and might depend on data. Performance of UIncp makes it the next best model, including reproducibility; however, it gets saturated after a certain threshold of a number of images in the training data set with less variation rendering its less attractive option. For data with large variation, UIncp performs equally with slightly low values but provides reproducible results. The performance of URsD is in-between to other models. UVgg is the best alternative for data with large variation if one does not want to estimate the threshold values required for URsEn. Its poor performance for data with less variation makes it a less attractive universal choice.

Following empirical expressions are presented to estimate the limits of models (for OCT data in hand):

Benign / untreated Tumor:

$$\text{Dice}= 0.729 + 0.004N_{Training} - 0.029N_{Testing}$$
$$\text{Loss}= -0.836 - 0.002N_{Training} - 0.012N_{Testing} \quad (11)$$

For Malignant / under treatment Tumor:

$$\text{Dice}= 0.272 - 0.005N_{Training} - 0.034N_{Testing}$$
$$\text{Loss}= -0.269 - 0.021N_{Training} - 0.145N_{Testing} \quad (12)$$

Eq. 11 and Eq. 12 indicate that the number of training and testing images in the data set both play a role in testing any particular model. The dependency of the number of images in the validation data set can also be explored in future work. Segmentation software must have several segmentation models incorporated with an analytical expression. The empirically obtained analytical expression would help selecting a particular model based on the type of tumor data and the number of images in the data.

This paper highlights a need for an add-on (in segmentation software) that would develop relation such as reported in Eqs. 11 and 12 to suggest (to medical professionals) aptness (for better accuracy, reproducibility, and speed) of available methods and performance indices regarding data before final processing takes place.

*We recommend similar sensitivity analysis should be included as standard practice before employing a particular deep learning model for biomedical image analysis.*

Declaration of Competing Interest
The author declares that he has no known competing financial interests or personal relationships that could have influenced the work reported in this paper.

CRediT authorship contribution statement
Mayank Goswami: Conceptualization, Methodology, Software, Investigation, Writing - original draft, Visualization, Supervision, Funding acquisition. Snehlata Shakya: Software, review & editing.

## Acknowledgments

This work is partially supported by GRANT Code I.M.P./2018/001045 by IMPRINT-II by S.E.R.B., Government of India. We would like to acknowledge the preliminary contribution made by Mr. Adityaojas Sharma, B.Tech, BT/IITR, and Mr. Kishan Kumar, B.Tech. EE/IITR. OCT Tumor Part of the Data was gathered at Eyepod, UCDavis, CA, U.S.A. by M.G. M.G. is grateful to the Alma Mater.